\documentclass{ws-ijmpcs}

\begin{document}

\markboth{M. Kuze}
{Energy-frontier lepton-hadron collisions: LHeC and FCC-eh}

%
\catchline{}{}{}{}{}
%

\title{Energy-Frontier Lepton-Hadron Collisions at CERN:\\
the LHeC and the FCC-eh}

\author{Masahiro Kuze}

\address{Department of Physics, Tokyo Institute of Technoogy\\
Tokyo, 152-8551,  Japan
\\
kuze@phys.titech.ac.jp}

\maketitle


\begin{abstract}
Lepton-hadron colliders that use a proton or nucleus beam of current and future hadron colliders and let it collide with an electron beam from a newly built electron accelerator bring attractive physics programs which are strong and complementary to the hadron collider physics.  Machine development for Energy Recovery LINAC and physics performance studies of such electron-hadron colliders, specifically the LHeC that uses the existing LHC beam and FCC-eh that is an option of Future Circular Collider program, are ongoing and reviewed in this article.
\keywords{Deep Inelastic Scattering; LHeC; FCC-eh.}
\end{abstract}

\section{Introduction}
Deep inelastic scattering (DIS) of electrons on protons or nuclei has been traditionally the best way to probe the inner structure of nucleon and nuclei.
At high energies, in addition to the electromagnetic (photon) exchange, the electroweak bosons play important roles; the $\gamma/Z$ exchange induces neutral current (NC) DIS, while the $W$ exchange results in charged current (CC) DIS, in which the outgoing neutrino is undetected and leaves a missing energy signature.
The scattering is described by two kinematic variables, $Q^2$, the squared momentum transfer between the lepton and hadron, and Bjorken $x$, the fraction of the nucleon momentum carried by the scattering parton.

The HERA collider at DESY was the last and highest-energy $ep$ collider which had a center-of-mass (cms) energy of 318~GeV.  It provided the PDF (parton distribution functions) of the proton up to the scale of $Q^2 \approx 10^4~\rm GeV^2$ and down to $x \approx 10^{-5}$, which are indispensable inputs to the physics at the LHC.

It is natural to consider the possibilities of future colliders using the proton (or heavy-ion) beam of a hadron collider and let it collide with an electron beam (polarizable) from a newly built electron machine.  Two ideas have been discussed; the LHeC\cite{CDR} collides a 60~GeV $e$-beam with the 7~TeV $p$-beam of the LHC, with a center-of-mass (cms) energy of 1.3~TeV, and FCC-eh collides a 60~GeV $e$-beam with the 50~TeV $p$-beam of the planned FCC (Future Circular Collider), with a cms energy of 3.5~TeV.  Both ideas have an option to use a beam of nuclei in addition to the proton beam.

Since such a facility uses a beam of the already built hadron collider, it can be realized at an affordable cost.
It can run concurrently with hadron-hadron collision experiments, and provides much cleaner collision environment than $h$-$h$ experiments (negligible pile-up), while realizing higher cms energy than $e^+e^-$ colliders.

\begin{figure}[htb]
\vspace{-0.3cm}
\centerline{\includegraphics[width=10cm]{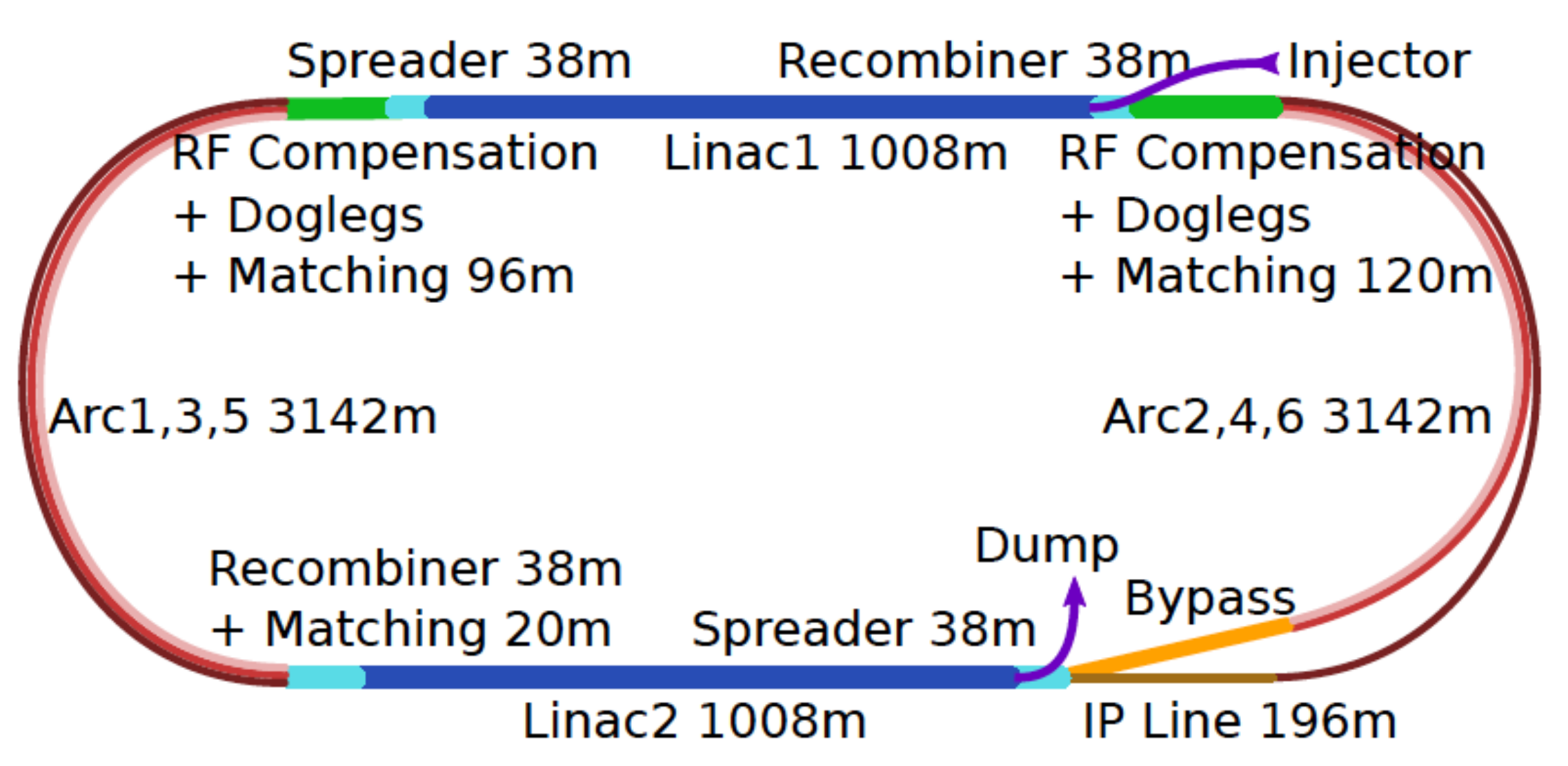}}
\vspace{-0.1cm}
\caption{Layout of Energy Recovery LINAC\protect\cite{fig1}.\label{f1}}
\vspace{-0.6cm}
\end{figure}

\section{Machine and Detector}	
The proposed electron machine for LHeC/FCC-eh is an energy recovery LINAC (ERL), which is a horserace-track like ring with two 10~GeV LINACs.  After three turns, the beam is accelerated to 60~GeV.
The circumference of the ring is approximately 9~km (see Fig.~\ref{f1}).
A unique characteristic of the ERL is that the beam after the collision runs in the same LINAC at an opposite phase to the accelerated beam and is thus {\it decelerated}, giving back the power for acceleration.  In this way the RF power is {\it recycled} and a lot of wall-plug power consumption can be saved.

The LINAC has a series of 802~MHz five-cell superconducting cavities with an accelerating gradient of 18~MV/m.
With high current electron beam, the collider aims at an instantaneous luminosity of $10^{34}~ {\rm cm^{-2} s^{-1}}$,
providing physics dataset of 100~$\rm fb^{-1}$ per year.
A small-scale ERL demonstrator called PERLE\cite{PERLE} is proposed at LAL, Orsay.
It will have two LINACs with four cavities each, which after three turns give $\approx$ 400~MeV beam of $\approx$ 15~mA.  The main purpose of PERLE is to probe the ERL operation in multi-Megawatt regime and the multipass mode with a very high current, but also a low-energy, high-intensity $ep/eA (\gamma p/\gamma A)$ physics program can be envisaged.

\begin{figure}[htb]
\vspace{-0.3cm}
\centerline{\includegraphics[width=10cm]{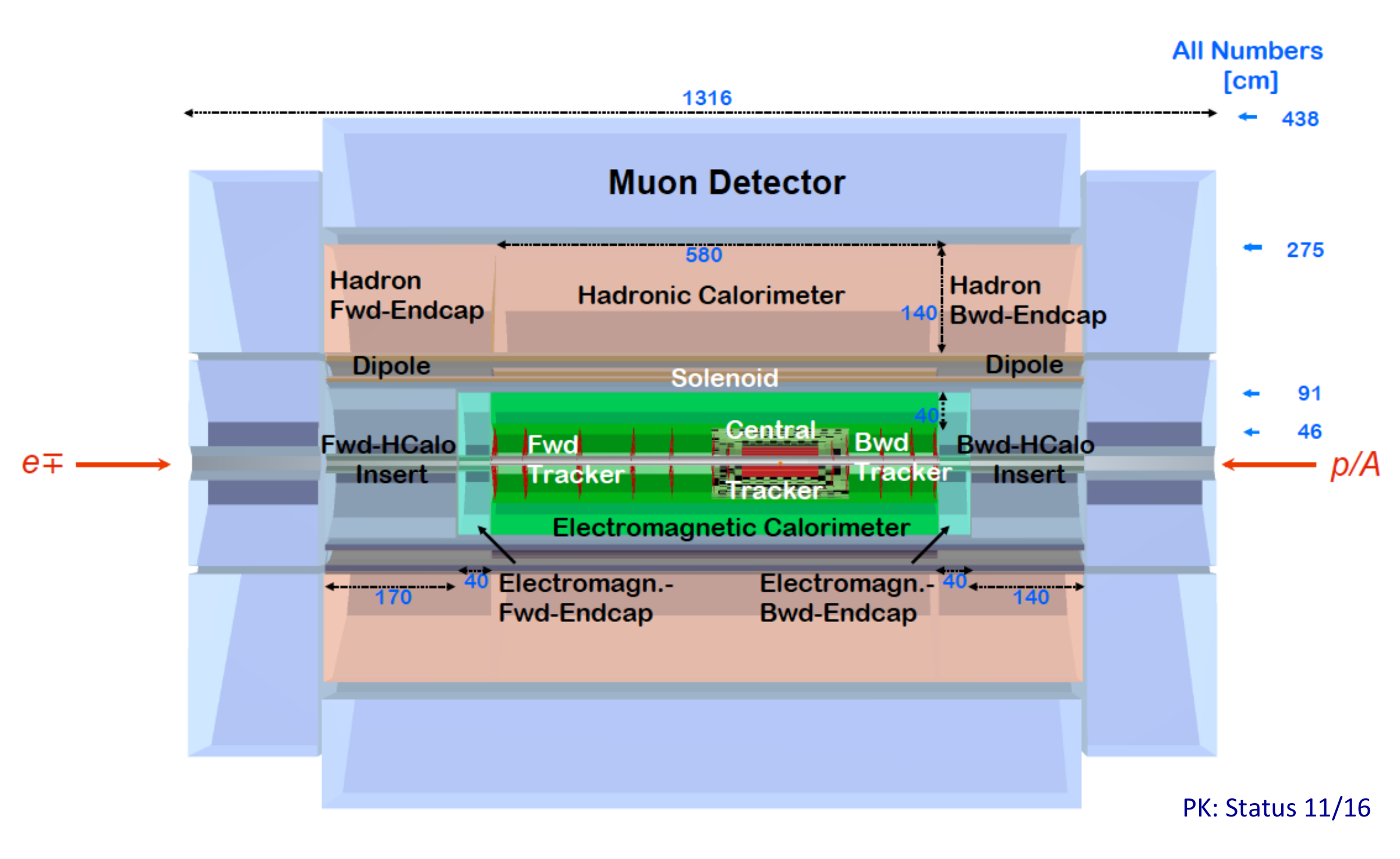}}
\vspace{-0.4cm}
\caption{Schematic of a future $ep$ detector\protect\cite{fig2}.\label{f2}}
\vspace{0.0cm}
\end{figure}

Also detector designs are ongoing in the LHeC/FCC-eh working group aiming at optimization of physics performance.
Because of the large asymmetry of the beam energies, the detector is also asymmetric like the detectors at HERA.  Very low-angle tagging of particles is important so the detector coverage extends to high repidity.
Fig.~\ref{f2} shows a schematic of a detector design.

\section{Physics Opportunities at Lepton-Hadron Colliders}
A variety of physics programs are discussed for LHeC/FCC-eh, which are unique and complementary to hadron collider physics programs.  Some highlights among them are briefly discussed in the following.

\subsection{Ultimate precision measurements of PDF and $\alpha_S$}
\begin{figure}[htb]
        \begin{minipage}[t]{0.45\linewidth}
        			\vspace{-0.5cm}
	                \hspace{0.5cm}
            \centerline{\includegraphics[width=5.5cm]{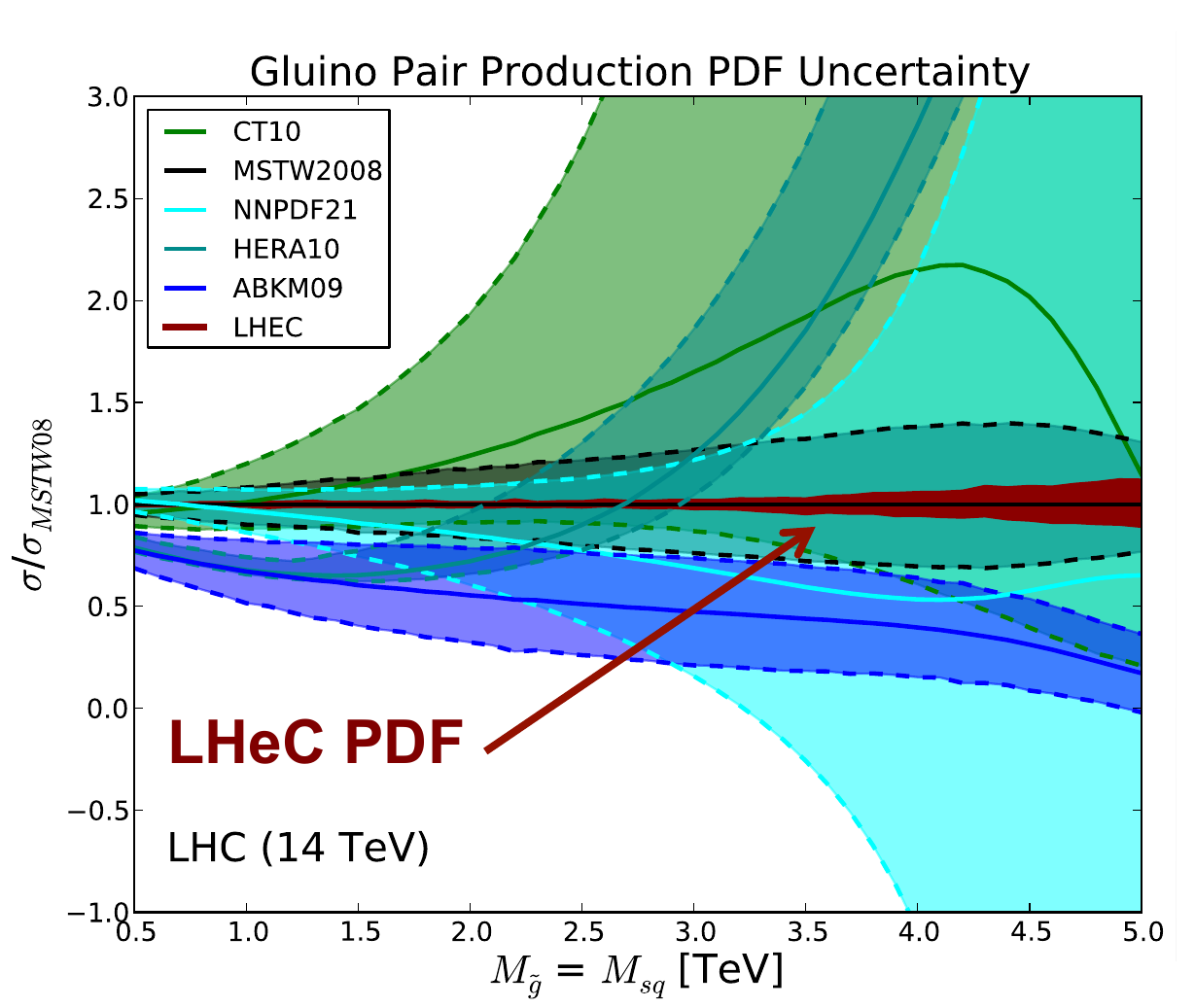}}
            \vspace{-0.4cm}
            \caption{Uncertainty of gluino pair production cross section at LHC\protect\cite{1211.5102}}
            \label{f3}
        \end{minipage}
        \hspace{0.1\linewidth}
        \begin{minipage}[t]{0.35\linewidth}
			\vspace{-0.3cm}
                        \hspace{8cm}
          \centerline{\includegraphics[width=6.5cm]{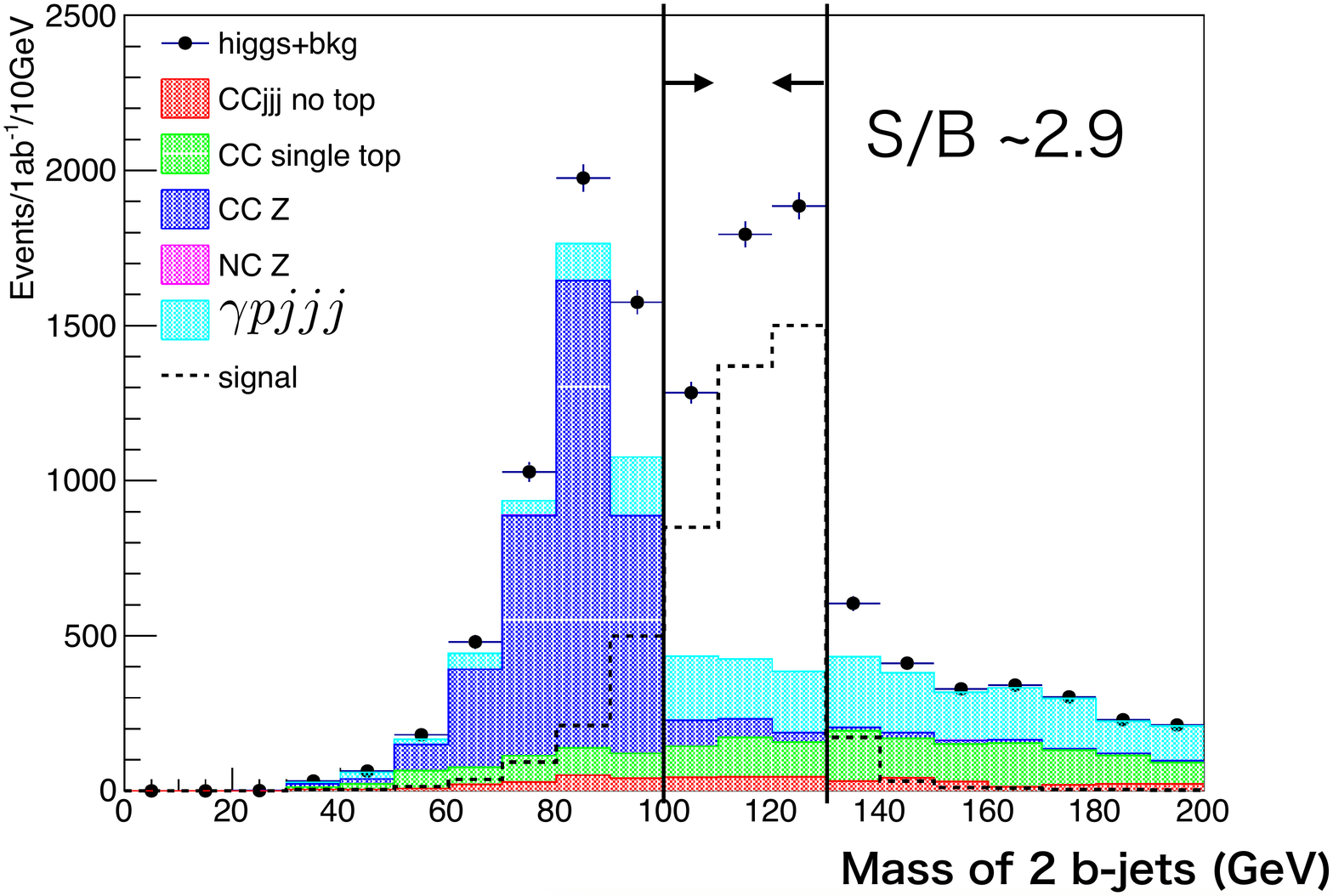}}
          \vspace{-0.55cm}
            \caption{Invariant mass of $H\to b \bar b$ signal on top of backgrounds\protect\cite{fig4}}
            \label{f4}
        \end{minipage}
    \end{figure}

It is clear from LHC experience that precise knowledge of PDFs is vital information for searches and precision physics at hadron colliders.  High-$x$ region is relevant for searches for new, very high mass particles.  At very large collision energy like FCC (cms energy of 100~TeV), small-$x$ region below $10^{-5}$ becomes relevant even for {\it common} physics targets such as $W/Z$, Higgs or top quarks.
Figure~\ref{f3} shows the uncertainty of gluino pair production cross section at the LHC (cms energy 14~TeV).
Using the current PDF sets on the market, the uncertainty exceeds 100\% above gluino mass of 2~TeV.
With the knowledge of precise PDF from LHeC measurements, this uncertainty squeezes to below 10\%.
Also the uncertainty of Higgs production cross section from PDF shrinks enough using the LHeC PDF so that the cross section measurement at the LHC becomes sensitive to the Higgs mass value\cite{1305.2090}.

From the DIS measurements, also the strong coupling constant $\alpha_S$ can be extracted by a simultaneous fit with PDFs.  A precision of 0.3\% is expected at the LHeC, which can further improve to 0.1\% when combined with HERA results\cite{Gwenlan}.

\subsection{$ep$ collider as a Higgs factory}
The production cross section of Higgs bosons in $ep$ collision lies at sub-picobarn range at LHeC and FCC-eh energies, which makes it very attracting for precise Higgs studies.
The CC channel is particularly interesting due to suppression of NC DIS background and the possibility of increasing the cross section with electron polarization ($-80\%$ is assumed in the performance evaluation).

Figure~\ref{f4} shows an invariant mass distribution for $H\to b \bar b$ reconstruction with a cut-based analysis assuming 10 years (1000~fb$^{-1}$) of data taking.  In the signal mass window, 3600 signal events are observed on top of 1250 background events.  This leads to a $Hbb$ coupling measurement of below 1\% precision (statistical error only).

Using multivariate techniques, much developed in the LHC physics, one can further improve the precision and could even measure the $Hcc$ coupling which is deemed very difficult at hadron colliders.
Table~\ref{ta1} summarizes the precision of the couplings anticipated at several configurations including the one using Double-energy LHC (proton beam energy of 14~TeV with stronger magnets in the existing LHC tunnel).

\begin{table}[ph]
\tbl{Higgs coupling precisions expected at future lepton-hadron colliders\protect\cite{Uta}.
$E_e$=60~GeV is assumed.}
{\begin{tabular}{@{}cccc@{}} \toprule
 & LHeC & $e$+DLHC  & FCC-eh \\
 & ($E_p$=7~TeV, & ($E_p$=14~TeV, & ($E_p$=50~TeV, \\
Coupling & $\sqrt{s}\approx$1.3~TeV) & $\sqrt{s}\approx$1.8~TeV) & $\sqrt{s}\approx$3.5~TeV)  \\ \colrule
$\kappa(Hbb)$ & 0.5\% & 0.3\% & 0.2\% \\
$\kappa(Hcc)$ & 4\%& 2.8\% & 1.8\% \\ \botrule
\end{tabular} \label{ta1}}
\vspace{-0.5cm}
\end{table}

\subsection{Top and electroweak physics}
    \begin{figure}[htb]
        \begin{minipage}[t]{0.4\linewidth}
        			\vspace{-0.1cm}
	                \hspace{0.2cm}
            \centerline{\includegraphics[width=4.2cm]{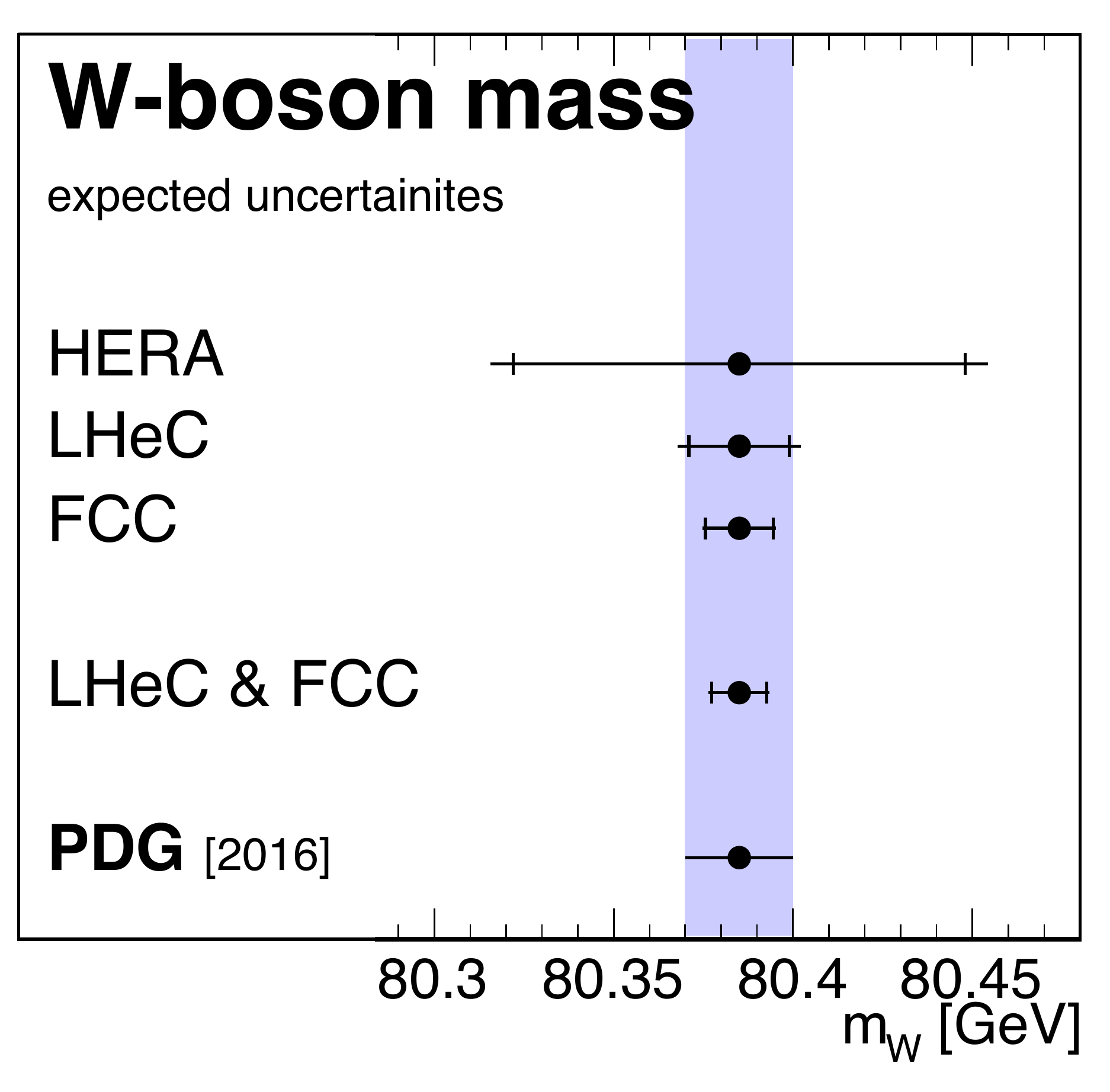}}
            \vspace{-0.4cm}
            \caption{$W$ mass measurement accuracy at HERA and future $ep$ colliders\protect\cite{fig5}.  As a reference, current accuracy from ATLAS\protect\cite{ATLASW} is 19~MeV, slightly larger than the PDG2016 band ($\pm 15$~MeV).}
            \label{f5}
        \end{minipage}
        \hspace{0.09\linewidth}
        \begin{minipage}[t]{0.35\linewidth}
			\vspace{0.3cm}
                        \hspace{-0.1cm}
          \centerline{\includegraphics[width=8cm]{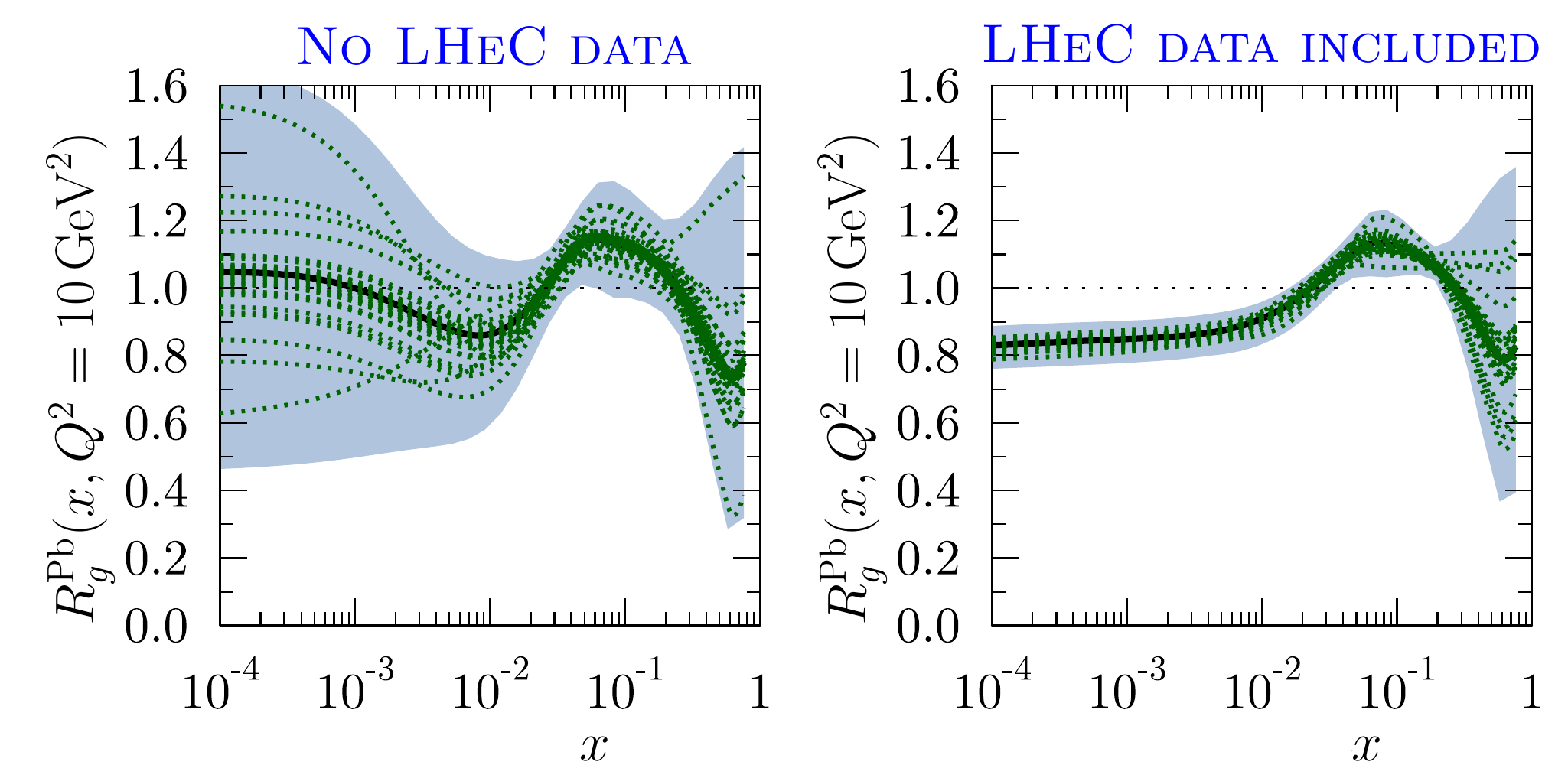}}
          \vspace{-0.6cm}
            \caption{Anticipated improvement on nuclear PDF precision at LHeC (gluon distribution in lead)\protect\cite{fig6}}
            \label{f6}
        \end{minipage}
    \end{figure}
The energy-frontier electron-hadron collider produces also a large number of top quarks and $W/Z$ bosons,
with little background from hadronic {\it QCD} events and pileup events inherent to hadron colliders.
Figure~\ref{f5} shows a prospect of $W$ mass measurements from CC DIS data from LHeC, FCC-eh and a combination of them.  A very competitive measurement can be made in comparison with the current accuracy.
The possibility of polarizing the electron beam (up to 80\% is expected) brings further opportunities in the electroweak measurements.
A search for single-top production brings a competitive test of FCNC top couplings, especially with $u$ quarks which are abundant in the proton.

\subsection{Beyond SM physics}
It is fair to say that the highest-energy hadron collider is the front runner in the discovery of new heavy particles or states,
but there are places where $ep$ collider can make a case.
An example is a leptoquark (LQ), a hypothetical state that couples directly with a lepton and a quark.
It is expected that LQs found in HL-LHC would be also found at $ep$ colliders.
Then, $ep$ colliders can study thoroughly the characteristics of the new particle, by determining its quantum numbers such as lepton/baryon numbers, spin and generation indices, thanks to the ability to control the electron beam charge and polarization\cite{BSM}.
Other topics of interest include compositeness, charged Higgs, sterile neutrinos, long-lived parties, or anomalous couplings.

\subsection{Diffractive physics and nuclear PDF}
Another interesting area to be probed is the low-$x$ and diffractive physics.
Compared to HERA, the reachable kinematics is much enhanced.
At very low-$x$ below $10^{-4}$, there is no data to constrain the gluon distribution,
which is expected to saturate somewhere.
Also a lot of diffractive measurements can be done, using rapidity gap events
or installing roman-pot type forward proton spectrometers\cite{diff}.

If a beam of nuclei is available in the hadron machine (like Pb in the LHC),
the first measurement of nuclear PDFs using electon-hadron collider can be made (note that HERA circulated only protons as the hadron beam).
Compared to the past measurements from fixed-target experiments,
the gain of kinematics is {\it four} orders of magnitude in $x$ and $Q^2$.
Figure~\ref{f6} shows an example of improvement in the accuracy of nuclear PDF measurement.

\section{Conclusions}
A new electron-hadron collider, using a hadron beam of existing or planned hadron colliders, is a cost-effective and attractive future program.
A design of an ERL with 60~GeV electron beam is at an advanced state, and a demonstrator PERLE is proposed.
An $ep/eA$ energy frontier machine, with 100 times $Q^2$ reach and 1000 times integrated luminosity compared to HERA, will bring a rich physics program which is complementary to, and strengthens, the discovery programs at HL-LHC and FCC-hh.
It has a different physics objectives from the Electron Ion Collider\cite{EIC}  (EIC) in US, which is a lower energy machine and focuses on spin and medium-$x$ structure of nucleon and nuclei.

\end{document}